\documentclass[12pt]{article}
\usepackage{amsxtra,amssymb,amsthm,amsmath,latexsym}

\theoremstyle{plain}

\def\oH{\buildrel\circ\over H}
\def\oH1{\buildrel\circ\over H\kern-.02in{}^1}
\def\qed{{\hfill $\Box$}}

\begin{document}

%\begin{titlepage}

\title{Analysis of the Newton-Sabatier scheme for inverting
fixed-energy phase shifts}
   %\thanks{key words: 
    %}
  % \thanks{Math subject 
%classification: 34R30; PACS:  03.80.+r. 03.65.Nk }

 %\thanks{This paper was written when the author was visiting
%Institute
%for Theoretical Physics, University of Giessen. The author thanks
%DAAD for support and Professor W.Scheid for discussions}

\author{ A.G. Ramm
 \\ Mathematics Department, Kansas State
University, \\
Manhattan, KS 66506-2602, USA\\
ramm@math.ksu.edu\\
}

\date{}

\maketitle\thispagestyle{empty}

\begin{abstract}

\footnote{Math subject classification: 34R30; PACS:
03.80.+r. 03.65.Nk }
\footnote{This paper was written when the author
was visiting Institute for Theoretical Physics, University of
Giessen. The author thanks DAAD for support and Professor W.Scheid for
discussions} \\

Suppose that the inverse scattering problem is understood
as follows: given fixed-energy phase shifts, corresponding to
an unknown potential $q=q(r)$ from a certain class, 
for example,  $q \in L_{1,1}$, recover this
potential. Then
it is proved that the Newton-Sabatier (NS) procedure 
does not solve the above problem. It
is not a
valid inversion method, in the following sense: 1) it is not possible
to carry this procedure
through for the phase shifts corresponding to a
generic potential $q \in L_{1,1}$, where
$L_{1,1} := \{ q : q= \overline q, \int^\infty_0 r |q(r)| dr < \infty 
\}$ and recover the original potential: 
the basic integral equation, introduced by R. Newton without
derivation, in general, may be not solvable for some $r>0$, and if it
is 
solvable for all $r>0$, then the resulting potential is not equal to
the original generic $q \in L_{1,1}$. Here a generic $q$ is any
 $q$ which is not a restriction to $(0, \infty)$ of an analytic
function. 
 2) the ansatz $(\ast)$
$K(r,s) = \sum^\infty_{l=0} c_l \varphi_l (r) u_l (s)$,
used by R. Newton, is incorrect:
the transformation operator $I-K$, corresponding to a generic $q \in
L_{1,1}$, does not have $K$ of the 
form $(\ast),$
and  3) the set of potentials $q \in L_{1,1},$ 
that can possibly be obtained by NS
procedure, is not dense in the set of all $L_{1,1}$ potentials in the norm
of $L_{1,1}$. Therefore one cannot justify NS procedure even for
approximate solution of the inverse scattering problem with
fixed-energy phase shifts as data.
Thus, the NS procedure, if considered as a method
for solving the above inverse scattering problem, is based on an
incorrect ansatz,
the basic integral equation of NS procedure
is, in general, not solvable for some $r>0$, and in this case
this procedure breaks down, and NS procedure is
not an inversion theory: it
cannot recover generic potentials $q \in L_{1,1}$
from their fixed-energy phase shifts.

Suppose now that one considers another problem: given 
 fixed-energy phase shifts, corresponding to some potential,
find a potential which generates the same phase shifts.
Then NS procedure does not solve this problem either: the basic
integral equation, in general, may be not solvable for some $r>0$, and
then NS procedure breaks down.  
\end{abstract}

%\end{titlepage}

\section{Introduction and conclusions}

The NS procedure is described in \cite{N} and \cite{CS}. In the
sixties P. Sabatier published several papers concerning this
procedure, and there are quite a few papers of several authors using
this procedure and generalizing it. A vast bibliography of this topic
is given in \cite{CS} and \cite{N}, and by this reason we do not
include references to many papers treating this topic.     

In our arguments below two cases are discussed.
The first case deals with the inverse scattering problem with 
fixed-energy phase shifts as the data. This problem is understood
as follows: an unknown spherically symmetric potential $q$ from an a
priori fixed class, say $L_{1,1}$, a standard scattering class,
generates fixed-energy phase shifts  $\delta_l, l=0,1,2, \dots,$.
The inverse scattering problem consists of recovery of  $q$
from these data.

The second case deals with a different problem: given some
numbers  $\delta_l, l=0,1,2, \dots,$, which are assumed to
be fixed-energy phase shifts of some potential  $q$, from a class not
specified, find some potential  $q_1$, which generates 
fixed-energy phase shifts equal to  $\delta_l, l=0,1,2, \dots,$.
This potential  $q_1$ may have no physical interest because
of its non-physical" behavior at infinity or other  undesirable
properties.                   
 
We first discuss NS procedure assuming that it is intended to solve
the inverse scattering problem in case 1. 
Then we discuss NS procedure assuming that it is intended
to solve the problem in case 2.

{\bf Discussion of case 1:}

In \cite{N62} and \cite{N} a procedure was proposed by R. Newton for
inverting fixed-energy phase shifts $\delta_l, l=0,1,2, \dots,$
corresponding to an unknown spherically symmetric potential $q(r)$.
R. Newton did not specify the class of potentials for which
he tried to develop an inversion theory and did not formulate 
and proved any results which would justify the inversion procedure
he proposed (NS procedure). 
His arguments are based on the following  claim N1, which
is implicit in his works, but crucial for the validity of
NS procedure: 

{\it Claim N1: the basic integral equation 
$$ K(r,s) = f(r,s) - \int^r_0 K(r,t) f(t,s) \frac{dt}{t^2}, \quad 
0\leq s
\leq r<\infty, \eqno{(1.1)}
$$  
is uniquely solvable for all $r>0$.} 

Here           
$$ f(r,s) := \sum^\infty_{l=0} c_l u_l (r) u_l (s), \quad
  u_l := \sqrt{\frac{\pi r}{2}} J_{l + \frac{1}{2}}(r),
  \eqno{(1.2)} $$ 
$ c_l$ are real numbers,
the energy $k^2$ is fixed: $k=1$ is taken without loss of generality,
$J_{l + \frac{1}{2}}(r)$ are the Bessel functions. If  equation
 (1.1) is uniquely solvable for  all $r>0$, then the potential $q_1,$
that NS procedure yields, is defined by the formula:
$$ 
q_1(r) = -\frac{2}{r} \frac{d}{dr} \frac{K(r,r)}{r}. \eqno{(1.3)}
$$  
   
 The R. Newton's ansatz (1.1)-(1.2) for the transformation
kernel $K(r,s)$ of the 
Schroedinger operator, corresponding to
some $q(r)$, namely, that $K(r,s)$ is the unique solution to
(1.1)-(1.2), is
not correct for a generic potential, as follows from 
our argument below (see the justification of Conclusions). 

{\it If for some $r>0$ equation (1.1) is not uniquely solvable, then NS
procedure breaks down: it leads to locally non-integrable potentials
for which the scattering theory is, in general, not available
(see \cite{R3} and [1] for a proof of the above statement) .}

In the original paper \cite{N62} and in his book
\cite{N} R. Newton did not 
study the question,  fundamental for any 
inversion theory: does the reconstructed potential $q_1$ generate the 
data from which it was reconstructed?

In \cite{CS},  p. 205, there are two claims:

i) that $q_1(r)$ generates the original shifts $\{\delta_l\}$
"provided
that $\{\delta_l\}$ are not "exceptional"",
and 
ii) that NS procedure "yields one (only one) potential which
decays faster than $r^{-\frac{3}{2}}$" and generates the original
phase shifts
$\{\delta_l\}$.   

If one considers NS procedure as a solution to inverse scattering
problem of finding an unknown potential $q$ from a certain class, for
example $q(r)\in L_{1,1} := \{ q : q= \overline q, \int^\infty_0 r
|q(r)|dr < \infty \}$, from the  fixed-energy phase shifts, generated
by this  $q$,
then 
the proof, given in \cite{CS}, of claim
i) is not convincing:
it is not clear why the potential $q_1$, obtained by NS procedure,
has the transformation operator generated by the potential
corresponding to 
 the original data, that
is, to the given fixed-energy phase shifts. In fact, as follows from
Proposition 1
below, the potential  $q_1$ cannot generate the
kernel $K(r,s)$ of the transformation operator corresponding to
a generic original potential  
$q(r)\in L_{1,1} := \{ q : q= \overline q, \int^\infty_0 r |q(r)|   
dr < \infty \}$. 

Claim ii) is incorrect because the original generic
potential $q(r)\in L_{1,1}$ generates the phase shifts
$\{\delta_l\},$ and if $q_1(r),$
the potential obtained by NS procedure and therefore not
equal to  $q(r)$ by Proposition 1,  generates the
same phase shifts
$\{\delta_l\}$, then one has two different potentials $q(r)$ and
$q_1(r)$, which
both decay faster than $r^{-\frac{3}{2}}$ and both generate the
original phase shifts $\{\delta_l\},$ contrary to claim ii).

The purpose of this paper is to formulate and justify the following

{\bf Conclusions:}

{\it Claim N1 and ansatz (1.1)-(1.2) are not proved by R.Newton and,
in general,  are wrong.
Moreover, one cannot approximate with a prescribed accuracy
in the norm $||q||:=\int_0^\infty r|q(r)|dr$ a
generic potential 
$q(r) \in L_{1,1}$
by the potentials which might possibly
be obtained by the NS procedure. Therefore NS procedure
cannot be justified even as an approximate inversion procedure.}

{\bf Let us justify these conclusions:}

 Claim N1, formulated above and basic for NS procedure,
is wrong, in general, for the following reason:

 Given fixed-energy phase shifts, corresponding to a generic potential
$q \in L_{1,1}$,
one either cannot carry through NS procedure because:

a) the system (12.2.5a) in \cite{CS}, which should determine numbers
$c_l$ in formula (1.2), given the phase shifts $\delta_l,$
may be not solvable, or

b) if the above system is solvable, equation (1.1)
 may be not (uniquely) solvable for some $r>0$, 
and in this case NS procedure breaks down since it
yields a potential which is not locally integrable (see [9]
for a proof).

 If equation (1.1) is solvable for all $r>0$ and yields
a potential $q_1$ by formula (1.3), then this potential is not equal
to the original generic potential $q \in L_{1,1}$, as follows from 
Proposition 1, which is proved in [9] (see also [1]):

{\bf Proposition 1. } {\it If equation (1.1) is solvable for
all $r>0$
and yields a potential $q_1$ by formula (1.3), then this  $q_1$
is a restriction to $(0, \infty) $ of a function analytic
in a neighborhood of $(0, \infty) $.}

Since a generic potential $q\in L_{1,1}$ is not
 a restriction to $(0, \infty) $ of an analytic function,
one concludes that even if equation (1.1) is solvable for
all $r>0$, the potential $q_1$, defined by formula (1.3),
is not equal to the original generic potential  $q\in L_{1,1}$
and therefore the inverse scattering problem
of finding an unknown  $q\in L_{1,1}$ from its fixed-energy phase
shifts is not solved by NS procedure.

The ansatz (1.1)-(1.2) for the transformation kernel is, in general,
incorrect, as follows also from Proposition 1.

Indeed, if the ansatz (1.1)-(1.2) would be true and formula (1.3) 
would yield the original generic $q$,
that is $q_1=q$, this would contradict Proposition 1.
If formula (1.3) would yield a $q_1$ which is different from the
original generic $q$, then NS procedure does not solve the inverse
scattering problem formulated above. Note also that
it is proved in [10] that independent of
the angular momenta $l$ transformation operator, corresponding to a
generic $q\in L_{1,1}$ does exist, is unique, and is defined by
a kernel $K(r,s)$ which 
cannot have representation (1.2), since it yields by the formula  
similar to (1.3) the original generic potential $q$, which is not
a restriction of an analytic in a neighborhood of $(0,\infty)$
function to $(0,\infty)$.

The conclusion, concerning impossibility of approximation
of a generic  $q \in L_{1,1}$ by potentials  $q_1$, which can possibly
be obtained by NS procedure, is proved in section 2, see proof of
Claim 1 there.

   Thus, our conclusions are justified. $\Box$

Let us give some additional comments concerning NS procedure. 

  Uniqueness of the solution to the inverse problem in case 1
was first proved by A.G.Ramm in 1987 (see \cite {R1}  and references
therein)  for a class of compactly supported potentials, while 
R. Newton's procedure was published in \cite {N62}, when no uniqueness
results 
for this inverse problem were known.
It is still an open problem if for the standard in scattering
theory class of $ L_{1,1}$ potentials the uniqueness theorem 
for the solution of the above inverse scattering problem holds. 

We discuss the 
inverse scattering problem with fixed-energy phase shifts (as 
the data) for potentials $q \in L_{1,1} $, because 
 only for this class of potentials a general theorem
of existence and
uniqueness of the transformation operators, independent of 
the angular momenta $l$, has been proved,  see \cite{R4}. 
 In \cite{N62}, \cite{N}, and in \cite{CS} this result was not
formulated and proved, and it was not clear for what class of
potentials 
the transformation operators, independent of $l$, do exist.
For slowly decaying potentials the existence of the transformation
operators, independent of $l$, is not established, in general, and
the potentials, discussed in [2] and [4] in connection with NS
procedure, are slowly decaying.

Starting with \cite{N62},  \cite{N}, and \cite{CS} claim N1
was not proved or the proofs given (see \cite{CT}) 
were incorrect (see \cite{R5}). This equation is uniquely
solvable for
sufficiently small $r>0$, but, in general, {\it it may be not
solvable  for some $r>0$, and if it is solvable for
all $r>0$, then it yields by formula (1.3)  a potential
 $q_1$, which is not equal
to the original generic potential $q\in L_{1,1}$, as follows from
 Proposition 1.}

Existence of "transparent" potentials is often cited in the literature.
A  "transparent" potential is a potential which is not equal to zero
identically, but generates the fixed-energy shifts which are all equal to
zero.

{\it In [2], p.207, there is a remark concerning the existence
of "transparent" potentials. This remark is not justified because it is
not proved
that for the values $c_l$, used in  [2], p.207, equation (1.1) is
solvable for all $r>0$. If it is not solvable even for one $r>0$,
then NS procedure breaks down and the existence of transparent
potentials is not established.}

 In the proof, given for the existence of the "transparent"
potentials in [2], p.197, formula (12.3.5), is used.
This formula involves a certain infinite matrix $M$. 
It is claimed in [2],  p.197, that this
matrix $M$ has the property $MM=I$, where $I$ is the unit matrix,
and on p.198, formula (12.3.10), it is claimed that a vector
$v \neq 0$ exists such that $Mv=0$. However, then $MMv=0$ and at the
same time $MMv=v \neq 0$, which is a contradiction. The difficulties
come from the claims about infinite matrices, which are not
formulated clearly: it is not clear in what space $M$, as an operator, 
acts, what is the domain of definition of $M$, and on what set of
vectors formula (12.3.5) holds. 

The construction of the "transparent" 
potential in [2] is based on the following logic: take all the fixed-energy
shifts equal to zero and find the corresponding $c_l$ from
the infinite linear algebraic system (12.2.7) in [2]; then
construct the kernel $f(r,s)$ by formula (1.2) and solve equation (1.1)
for all $r>0$; finally construct the  "transparent" potential by formula
(1.3). As was noted above, it is not proved that equation (1.1)
with the constructed above kernel  $f(r,s)$ is solvable for all
$r>0$. 
Thefore the existence of the
"transparent" potentials is not established. 
 
The physicists have been  using NS procedure without questioning
its validity for several decades.
Apparently the physicists still believe that NS procedure is
"an analog of the Gel'fand-Levitan method" for inverse
scattering problem with fixed-energy phase shifts as the data.
In this paper the author explains why such a belief is not
justified and why NS procedure is not a valid inversion method.
Since modifications of NS procedure are still used by
some physicists, who believe that this procedure
is an inversion theory, the author pointed out some
questions concerning this procedure in \cite{ARS} and \cite{R3},
and wrote this paper.

This concludes the discussion of case 1. $\Box$

{\bf Discussion of case 2:}

{\it Suppose now that one wants just to construct a 
potential $q_1$, which generates the phase shifts corresponding to
 some $q$.}

This problem is actually {\it not an inverse scattering problem} because
one does not recover an original potential from the scattering
data, but rather wants to construct some potential which
generates these data and may have no physical meaning.
Therefore this problem is much less interesting practically than
the inverse scattering problem.
 
{\it  However, NS procedure does not solve this
problem either: there is no guarantee that this procedure is
applicable, that is, that the steps a) and b), described 
in the justification of the conclusions,
can be done, in particular, that equation (1.1) is 
uniquely solvable for all $r>0$.}

If these steps can be done, then one needs to check that
the potential $q_1$, obtained by formula (1.3),  generates
the original phase shifts. This was not done in \cite{N62} and
\cite{N}.

This concludes the discussion of case 2. $\Box$

The rest of the paper contains formulation and proof
of Remark 1 and Claim 1.
  
It was mentioned in \cite{N67} that if $Q:=\int^\infty_0 r q(r) dr\neq
0,$ then the numbers $c_l$ in formula (1.2)  cannot 
satisfy the condition  $\sum_0^\infty
|c_l|<\infty$.   
This observation can be obtained also from the following 

{\bf Remark 1.} {\it For any potential $q(r) \in L_{1,1}$ such that
$Q:= \int^\infty_0 rq(r) dr \neq 0$ the basic equation
(1.1) is not solvable for some $r>0$ and any choice of $c_l$
such that 
$\sum^\infty_{l=0} |c_l| < \infty.$}

Since generically, for  $q \in L_{1,1},$ one has
 $Q\neq 0$, this gives an additional
illustration to the conclusion that equation (1.1),
in general, is not solvable for some $r>0$. 
Conditions $\sum^\infty_{l=0} |c_l| < \infty$ and $Q\neq 0$
are incompatible. 

In \cite{CS}, p. 196,
a weaker condition $\sum^\infty_{l=0} l^{-2} |c_l| < \infty$
is used, but in the
examples (\cite{CS} pp. 189-191), $c_l = 0$ for all $l \geq l_0 > 0$,
so that $\sum^\infty_{l=0} |c_l| < \infty$ in all of these examples.  

 {\bf Claim 1:}  {\it The set of the potentials $v(r)\in L_{1,1},$
which can possibly be obtained by the NS procedure,
is not dense (in the norm $\| q \| := \int^\infty_0 r |q(r)| dr$)
in the set $L_{1,1}$.}

In section 2 proofs are given.

\section{Proofs} %2

{\bf Proof of Remark 1:}
Writing (1.3) as
$K(r,r) = -\frac{r}{2} \int^r_0 sq_1(s) ds $
and assuming $Q \neq 0,$
one gets the following relation:
$$K(r,r) = -\frac{Qr}{2} \left[ 1+ o(1) \right] \to \infty \hbox{\ as\ }
  r \to \infty. \eqno{(2.1)}$$

If (1.1) is solvable for all $r>0$, then from (1.2) and (1.1) 
it follows that
$K(r,s) = \sum^\infty_{l=0} c_l \varphi_l (r) u_l (s),$
where
$\varphi_l (r) := u_l (r) - \int^r_0 K(r,t) u_l (t) \frac{dt}{t^2},
  $
so that $I - K$ is a transformation operator,
where $K$ is the operator with kernel $K(r,s)$,
$\varphi^{\prime \prime}_l + \varphi_l - \frac{l(l+1)}{r^2}
  \varphi_l - q(r) \varphi_l = 0, $
$q(r)$ is given by (1.4), $\varphi_l  = O(r^{l+1})$, as $r \to 0,$
$$u_l (r) \sim \sin \left(r - \frac{l \pi}{2} \right), \quad
  \varphi_l (r) \sim |F_l|\sin \left( r-\frac{l \pi}{2} + \delta_l
\right)
  \hbox{\ as\ } r \to \infty, $$
where $\delta_l$ are the phase shifts at $k=1$
and $F_l$ is the Jost function at $k=1$. 
It can be proved that $\sup_l |F_l|<\infty$. Thus,
if $\sum^\infty_{l=0} |c_l| <\infty,$ then
$$K(r,r) = O(1) \hbox{\ as\ } r \to \infty. \eqno{(2.2)}$$

If $Q\neq 0$ then (2.2) contradicts (2.1). It follows that if $Q \neq
0$ then equation
(1.1) cannot be uniquely solvable for all $r>0$, so that NS procedure
cannot be carried through if $Q \neq 0$ and $\sum^\infty_{l=0}
|c_l|<\infty.$ 
This proves Remark 1. \qed

{\bf Proof of Claim 1:}
Suppose that $v(r)\in L_{1,1}$ and $Q_v :=
\int_0^\infty rv(r) dr = 0$,
because otherwise NS procedure cannot be carried through as was
proved in Remark 1.

If $Q_v=0$, then there is also no guarantee that  NS procedure can
be
carried
through. However, we claim that if one assumes that it 
can be carried through,
then the set of potentials, which can possibly be 
obtained by NS procedure,
is not dense in $L_{1,1}$ in the norm
$\| q \| := \int^\infty_0 r |q(r)| dr$. In fact, any potential $q$ such
that $Q:= \int^\infty_0 r q(r) dr \neq 0,$
and the set of such potentials is dense in  $L_{1,1}$, cannot be
approximated with a prescribed accuracy by
the potentials which can be possibly obtained by the NS procedure.

Let us prove this. Suppose that $q \in L_{1,1}$,
$$Q_q := \int^\infty_0 rq(r) dr \neq 0, \hbox{\ and\ }
  \|v_n - q \| \to 0 \hbox{\ as\ } n \to \infty,$$
where the potentials $v_n \in L_{1,1}$ are obtained by the NS procedure,
so that
$$Q_n := \int^\infty_0 rv_n (r) dr = 0.$$
We assume $v_n \in L_{1,1}$ because otherwise $v_n$ obviously cannot
converge in the norm $||\cdot||$ to $q\in L_{1,1}$. 
Define a linear bounded on $L_{1,1}$ functional
$$f(q) := \int^\infty_0 rq(r) dr, \quad |f(q)| \leq \| q \|,$$
 where
$\| q \| := \int^\infty_0 r|q(r)| dr$. The potentials
$v \in L_{1,1}$, which can possibly be obtained by the NS procedure, belong
to the null-space of $f$, that is $f(v) = 0.$ 

If $\lim_{n \to \infty} \| v_n-q \| = 0$, then
$\lim_{n \to \infty} |f(q-v_n)| \leq \lim_{n \to \infty}
  \| q-v_n \| = 0. $
Since $f$ is a linear bounded functional and $ f(v_n)=0$, one gets: 
$f(q-v_n) = f(q) - f(v_n) = f(q)$. So if $f(q) \neq 0$ then
$$\lim_{n \to \infty} |f(q-v_n)| = |f(q)| \neq 0. $$
Therefore, no potential $q \in L_{1,1}$ with $Q_q \neq 0$ can be
approximated
arbitrarily accurately by a potential $v(r)\in L_{1,1}$ which can possibly
be
obtained by the NS
procedure. Claim 1 is proved. \qed

\end{document}